\documentclass[journal]{vgtc}                

\ifpdf
  \pdfoutput=1\relax                   
  \usepackage{graphicx}                
  \DeclareGraphicsExtensions{.pdf,.png,.jpg,.jpeg} 
\else
  \usepackage{graphicx}                
  \DeclareGraphicsExtensions{.eps}     
\fi%

\graphicspath{{figures/}{pictures/}{images/}{./}} 

\usepackage{microtype}                 
\PassOptionsToPackage{warn}{textcomp}  
\usepackage{textcomp}                  
\usepackage{mathptmx}                  
\usepackage{times}                     
\usepackage{cite}                      
\usepackage{tabu}                      
\usepackage{booktabs}                  

\onlineid{1014}

\vgtccategory{Research}
\vgtcpapertype{Algorithm/Technique}

\title{Deadeye: A Novel Preattentive Visualization Technique Based on Dichoptic Presentation}

\author{Andrey Krekhov, \textit{Student Member, IEEE}, and Jens Kr\"uger, \textit{Member, IEEE}}
\authorfooter{
\item
 Andrey Krekhov and Jens Kr\"uger are with Center of Visual Data Analysis and Computer Graphics (COVIDAG), University of Duisburg-Essen. E-mail: \{andrey.krekhov, jens.krueger\}@uni-due.de.
\item
 Jens Kr\"uger is also with SCI Institute, University of Utah. E-mail: jens@sci.utah.edu.
}

\usepackage{microtype}

\usepackage{ulem}
\makeatletter
\def\ifEmpty#1{\def\@temp{#1}\ifx\@temp\@empty}
\makeatother

\usepackage{xspace}

\usepackage{amsmath,amsfonts,amssymb}
\usepackage{url} 

\newcommand{\FG}[1]{Figure~\ref{#1}}

\newcommand{\shah}{{\textstyle \amalg{\kern-4.pt\amalg}}}

\shortauthortitle{Krekhov \MakeLowercase{\textit{et al.}}: Deadeye: A Novel Preattentive Visual Feature Based on Dichoptic Presentation}

\abstract{Preattentive visual features such as hue or flickering can effectively draw attention to an object of interest -- for instance, an important feature in a scientific visualization. These features appear to pop out and can be recognized by our visual system,  independently from the number of distractors. Most cues do not take advantage of the fact that most humans have two eyes. In cases where binocular vision is applied, it is almost exclusively used to convey depth by exposing stereo pairs. We present \textit{Deadeye}, a novel preattentive visualization technique based on presenting different stimuli to each eye. The target object is rendered for one eye only and is instantly detected by our visual system. In contrast to existing cues, Deadeye does not modify any visual properties of the target and, thus, is particularly suited for visualization applications. Our evaluation confirms that Deadeye is indeed perceived preattentively. We also explore a conjunction search based on our technique and show that, in contrast to 3D depth, the task cannot be processed in parallel. 
} 


\teaser{
  \centering
  \includegraphics[width=\linewidth]{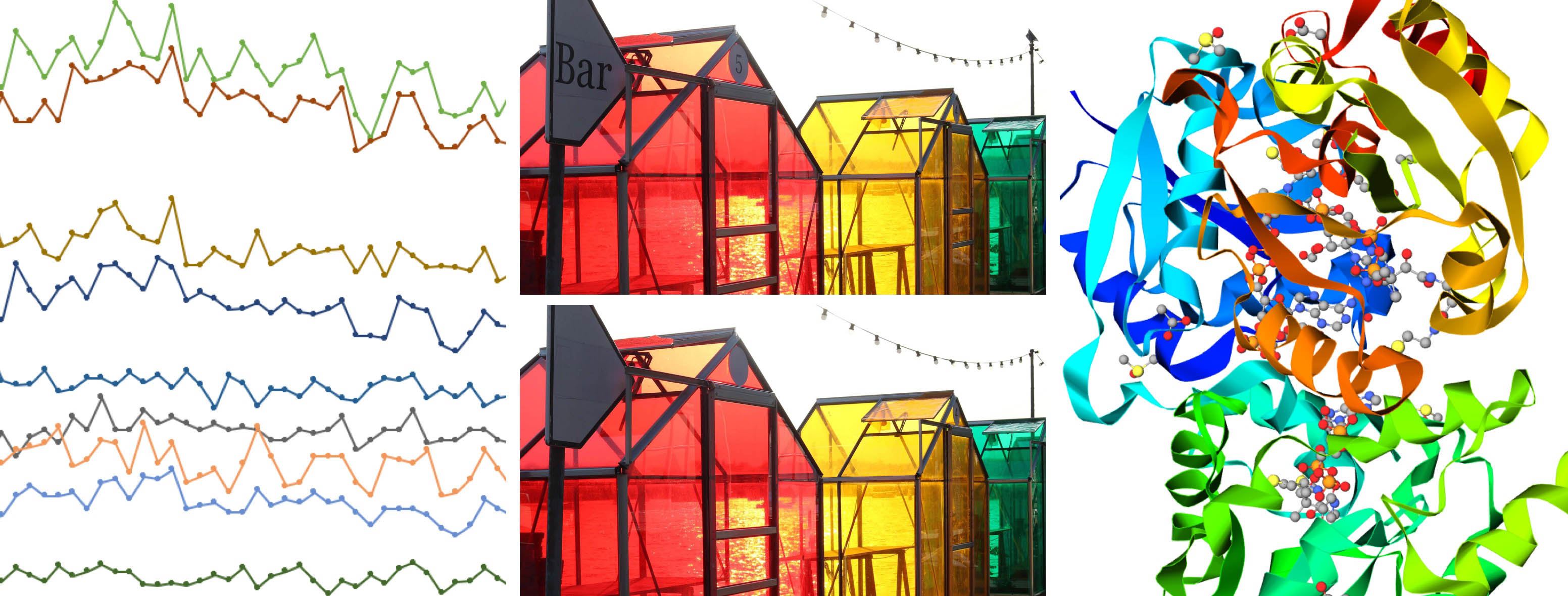}
  \caption{Deadeye is a preattentive visualization technique that allows to guide attention in various visualization scenarios. In contrast to existing methods such as flickering, shape, or color, our technique does not modify any visual property of the target object. Possible application scenarios include, e.g., highlighting of lines in a line chart (left), comparative visualization including its entertainment derivatives (middle: spot-the-difference puzzle), and scientific visualization of chemical reactions (right).}
	\label{fig:teaser}
}

\vgtcinsertpkg

\begin{document}


\firstsection{Introduction}

\maketitle


\begin{figure*}[t!]
\centering
\includegraphics[width=2.1\columnwidth]{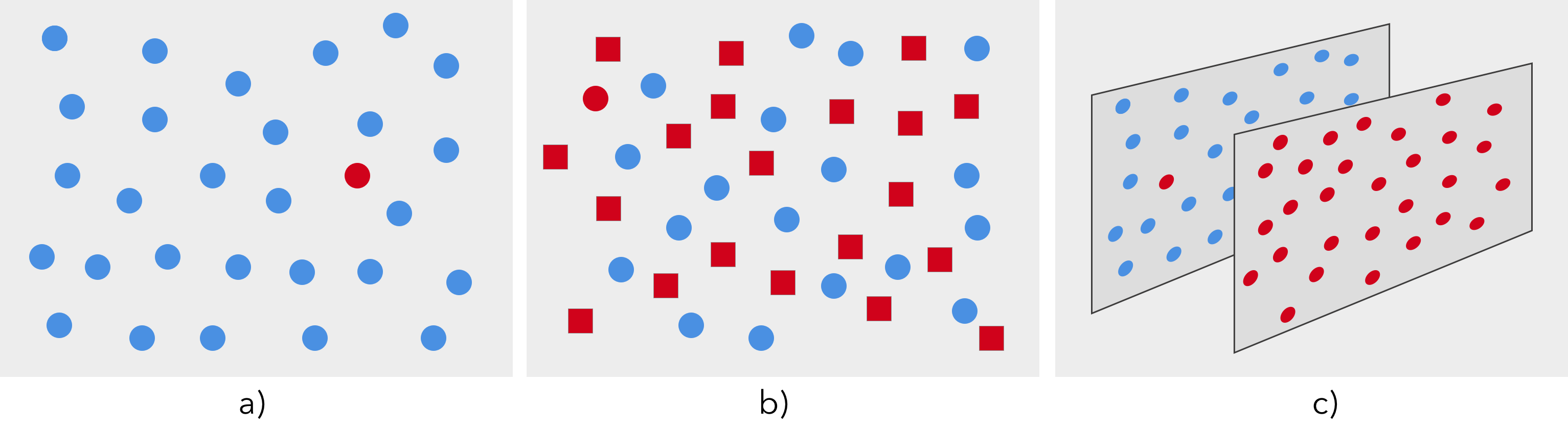}
\caption{Preattentive visual cues and conjunction search. (a) Color as cue: The target object is a red circle among blue distractors and can be recognized preattentively. (b) Conjunction of color and shape: The target object is either a blue square or a red circle. We have to search each object in a seral fashion to find the target. (c) Conjunction of stereo and color: The target object is either in the back plane and red or in the front plane and blue. Images redrawn from \protect\cite{Healey:2012:AVM:2225054.2225226} and \protect\cite{nakayama1986serial}.}
\label{fig:popout}
\end{figure*}

Designing comprehensive visualizations requires a deep understanding of how perception actually works. Therefore, the visual system is one of our most important tools for acquiring and parsing information that surrounds us. Making efficient use of certain visual characteristics helps us to create visualizations that excel in their usability and user performance. 

In particular, drawing the attention of users to certain elements is a research subject that is continuously being worked on in various fields, including psychology, computer science, psychophysics, and biology. Researchers have discovered several visual cues that can capture and guide our attendance to the object(s) of interest. The most prominent example is probably the search function of a PDF viewer, web browser, or text editor: it uses color to highlight the occurrence of a query, which allows us to instantly locate the results. A more sophisticated example is a medical visualization that utilizes flickering to  highlight suspicious cells or tissue and helps doctors with exploring the data. 

Cues such as color, flickering, shape, size, and motion are examples of so-called preattentive visual features. Such features make an object pop out and allow us to grasp the object's presence in usually less than 250 ms. One important property of preattentive cues is that they perform equally well with an increasing number of distractors -- the search time for our visual system remains constant. In other words, those features are processed in parallel by our visual system and are not searched in a serial fashion. That property is crucial when we revise our example with a full-page text search or the exploration of a huge medical dataset.

This paper summarizes \textit{Deadeye}, a novel preattentive visualization technique that is based on the dichoptic presentation phenomenon. We render two different images for the two eyes. The images have one difference. The object that should pop out is presented to only one eye. Surprisingly, previous research stated that binocular rivalry, an effect based on showing two different images, is not processed in parallel, with the exception of luster effects. Since then, those dichoptic techniques have lost popularity as possible popout effects. 

Our contribution is twofold. First, our finding opens the way for further research on that technique and encourages derived applications in visualization (cf. \FG{fig:teaser}). Second, Deadeye is the first preattentive technique that does not modify \textbf{any} visual properties of the target object. All existing cues have to alter the target in one way or another -- be it reshaping, recoloring or introducing motion. These changes can result in data misinterpretation, usually a highly undesired side effect. Furthermore, especially in visualization, most properties have a certain purpose and meaning, and reserving a whole dimension such as color or position for the visual popout is an expensive tradeoff. On the contrary, Deadeye solves these issues by preserving all visual properties of the target.

The paper shows that Deadeye is indeed perceived preattentively by conducting a state-of-the-art study that is common for such cues. Our evaluation also underlines the general applicability of the technique, as not a single participant reported headache or other similar physical strains. Three additional explorative experiments illustrate the usage of our technique in different real-world visualization scenarios.

In addition, we examine the phenomenon of a conjunction search based on Deadeye. Although preattentive features are processed in parallel, a conjunction of multiple cues usually leads to a serial search (cf. \FG{fig:popout}). In most cases, we have to inspect each object individually and check for its properties. Hence, the time needed for the task increases linearly with the number of distractors. However, there exist a few exceptions, one of which is 3D depth. Previous research shows that the depth cue can be combined with, e.g., color or motion, and still be processed in parallel. As our technique also makes use of the binocular visual system, one might assume that Deadeye also shares that property. To investigate that claim, we conduct a study by combining our technique with hue variations. The results clearly show a significant increase in time needed to accomplish the task and the overall accuracy, exposing the serial character of that task.

\section{Background and Related Work}

\subsection{Preattentive Visual Features}

The human visual system works in a rather dynamic way. Our eyes are able to perceive detailed information in a very limited area only. That area is determined by the fovea, the central part of our retina. The fovea provides high-resolution images but covers only between one and two degrees of vision. In order to gather all the data, our eyes perform rapid saccadic movements between the states of fixed steadiness, i.e., our eyes scan for interesting objects and determine the next location for gathering a high-resolution fragment. Humans are not aware of the process, since the interchange between saccades and fixations happens three to four times in a second. For a deeper insight into the basics of the human vision, please refer to the research by Yanbus~\cite{Yarbus1967, yarbus1967eye}, Noton et al.~\cite{NOTON1971929} and Itti et al.~\cite{itti2001computational}.

Early research on the interplay between eye movement and data processing has discovered several visual features that can be detected in a split second. That speed led to the assumption that these features can be detected preattentively. As pointed out by Healey et al.~\cite{Healey:2012:AVM:2225054.2225226}, the term \textit{preattentive} is not entirely correct, since a brief period of focus is still required to perceive these cues. However, we will stick to that term as it is used throughout the literature.

Such preattentive visual features can be detected within one focus period before the saccadic eye movement is triggered. Since the saccade has an initiation time of 200-250 ms~\cite{Healey:2012:AVM:2225054.2225226}, researchers apply that time threshold to verify if a visual variable is preattentive. In general, experiments are conducted as follows: an image with a number of distractors and possibly a target object is exposed to the participants, as shown in \FG{fig:popout}. The participants have to decide whether a target object was present. The image is shown for either less than 250 ms (e.g., \cite{Gutwin:2017:PPI:3025453.3025984}), or until the participants make the decision via a button press (e.g., \cite{vissearch}). In the first case, the error rate is considered as the primary indicator, whereas in the second case, one takes both the error rate and the reaction time into count. The experiments are conducted with varying set sizes in order to prove that the error rate and/or reaction time remain constant with an increasing number of distractor objects.

The work by Healey et al.~\cite{Healey:2012:AVM:2225054.2225226} summarizes and provides references for the following 16 preattentive visual cues: orientation, length, closure, size, curvature, density, number, hue, luminance, intersections, terminators, 3D depth, flicker, direction of motion, velocity of motion, and lighting direction. A prominent example is hue: consider a red dot that pops out in a set of blue distractor dots. For instance, Nagy et al.~\cite{nagy1990critical} conducted experiments to show how differences in color between distractors and the target object affect our search efficiency. As expected, a rather small color difference leads to an increased search time and diminishes our ability to process that visual feature in a preattentive way.

A more recent summary suited for a broader audience is exposed by Wolfe et al.~\cite{wolfe2017five}. Amongst other things, the article includes a classification of discovered features based on their likeliness to perform as guidance attributes. In addition, the authors summarize how the prior history of an observer influences the guidance of attention.

The most related visual feature to our research is the luster effect discovered by Wolfe et al.~\cite{wolfe1988binocularity}. Hereby, the target is rendered dimmer than the background for one eye and brighter than the background for the other eye. The result is often said to pop out because of a metallic sheen. The authors also experimented with other dichoptic presentation techniques and concluded that they cannot be perceived preattentively and that luster is an exception. Our paper partially supports these assumptions, as Deadeye also works preattentively. 

Another binocularity-based and closely related visual feature is 3D depth. Although previous research suggested that preattentive detection is solely based on 2D image features, Enns et al.~\cite{enns1990influence, enns1990sensitivity} found that our visual system can also access visual features related to the 3D scene information. The authors concluded that both the lighting direction and three-dimensionality can be detected preattentively. 


Different theories and models aim to explain the underlying nature of preattentive processing. Among the prominent ones is the feature integration theory by Treisman ~\cite{treisman1980feature}. A detailed description of such models is behind the scope of our paper. We point to the work by Healey et al.~\cite{Healey:2012:AVM:2225054.2225226} for a detailed state-of-the-art summary.

The model decomposes the low-level visual system in feature maps for each specific visual cue. Each map is encoded in parallel, which leads to an almost instantaneous detection of the feature. Other examples of such models include the texton theory~\cite{julesz1981textons} and the Boolean map theory~\cite{huang2007boolean}.

Clearly, each variable has its advantages and drawbacks. Certain features, such as shape or 3D depth, have to alter the underlying visualization and can lead to misinterpretation. Other features suffer in terms of accuracy when the target object is located in the peripheral area. A detailed study was conducted by Gutwin et al.~\cite{Gutwin:2017:PPI:3025453.3025984} to determine how different variables behave with varying distance from the focus point. This is a rather critical attribute, as multi-monitor setups gain more and more popularity. For these use-cases, Gutwin et al. suggest applying motion or flickering, as these variables perform well even at large angles. In the process, the authors applied the NASA-TLX survey, establishing comparisons among several variables. Therefore, we decided to rely on the same questionnaire to compare our outcome to the well-established visual cues.

\subsubsection{Application Areas}

Preattentive visual cues play an important role in visualization (e.g., \cite{ware2012information}) and other areas such as human-computer interaction (e.g., \cite{malamed2009visual}). One prominent use-case is directing the attraction of users to a certain object of interest. Several emphasis methods have recently been summarized by Hall et al.~\cite{Hall:2016:FEI} and methods for attention modeling in general can be found in the state-of-the-art report by Borji et al.~\cite{Borji:2013:SAV} and in the summary paper by Healey et al.~\cite{Healey:2012:AVM:2225054.2225226}. Not surprisingly, none of these summaries mention binocular disparities as a means for guiding attention or annotating objects. 

In the following paragraphs, we present a brief selection of established attention-guidance techniques. Hoffmann et al.~\cite{Hoffmann:2008:EVC:1357054.1357199} utilized visual cues in order to highlight the active window in a multi-monitor setup. The authors evaluated five types of window frames and masks and finally suggested a combination of color and tapered trails to guide the users' attention. Cole et al.~\cite{Cole:2006:DGM:2383894.2383942} introduced the \textit{Stylized Focus}, a variation of shading effects in order to draw the viewers' attention to a specific 3D object. The \textit{Popout Prism} presented by Suh et al.~\cite{Suh:2002:PPA:503376.503422} is an overview+detail-based system that enhances the representation of documents. The authors used the two popout effects color and size in order to visually emphasize elements of interest and also reduce the cognitive load. 

Alper et al.~\cite{Alper:2011:SHG:2068462.2068634} utilized the stereoscopic depth cue for highlighting purposes in 2D and 3D graph visualizations by projecting objects of interest onto a plane closer to the user. Additionally, the authors introduced a juxtaposing mechanism to allow focus and context views. The authors also emphasized the benefits of using highlighing mechanisms that do not reserve important visual attributes such as motion or color.

Possible applications and design guidelines for rather dynamic popout techniques flicker, direction, and velocity were studied by Huber et al.~\cite{huber2005visualizing}. The adoption of flickering for dynamic narrative visualizations, the so-called \textit{Attractive Flicker}, was discussed in detail by Waldner et al.~\cite{Waldner:2014:AFG}. In a first stage, user attention is attracted by a short and intensive flickering of the target object. The engagement stage relies on a less disturbing luminance oscillation and helps to keep track of the target. In contrast to cues such as color and size, the flickering does not distort the scene elements. Our technique goes a step further, since we do not alter any visible property of the object of interest.

%

\subsection{Dichoptic Presentation}

The term dichoptic presentation is applied when each eye is exposed to different stimuli, i.e., two different images. One of the resulting phenomena is called binocular rivalry (e.g., \cite{logothetis1996rivalling, blake1989neural, friedenberg2012visual, alais2005binocular, Paffen2011}). Instead of a stable single image, our vision switches to a mode with alternating periods of monocular dominance. One explanation is that the monocular neurons compete in the primary visual cortex and lead to the mentioned rivalry with regard to the interpretation of the image. Interestingly, as shown by Logothetis et al.~\cite{logothetis1996rivalling}, the rivalry does not depend on the eye, i.e., the dominant eye does not behave differently. One reason is that, in most cases, we are not able to determine which eye has detected a certain distinct stimulus. The research by Baker~\cite{baker2017decoding} addresses that possible loss of information by utilizing EEG and multivariate pattern analysis. Baker concludes that such eye-of-origin knowledge is lost due to our perception and consciousness. We also evaluate our technique for both eyes separately to explore the interplay between factors such as eye dominance and the preattentive processing of Deadeye.

Wolfe et al.~\cite{wolfe1988binocularity} claimed that binocular rivalry cannot be processed preattentively. The only exception is the luster effect that we mentioned previously. Zou et al.~\cite{zou2017binocularity} revisited that research and concluded that though interocular differences can guide attention in some cases (e.g., luster), the effects are rather weak and overridden by stronger features such as orientation or luminance. However, we show that Deadeye is another valid exception and propose to reconsider binocular rivalry in terms of its preattentiveness. Similarly to our suggestion, Friedenberg~\cite{friedenberg2012visual} also pointed out that it is still not clear whether binocular rivalry falls under late-stage voluntary attentional control or is processed by preattentive mechanisms. Another evidence to reconsider rivalry for attention guidance can be found in the work of Paffen et al.~\cite{paffen2012interocular}. The authors conducted a study with ten participants to see how transparent, monocular, and binocular changes in images are perceived. They concluded that change blindness is attenuated in cases where the change is monocular. The work by Zhaoping~\cite{zhaoping2008attention} also aligns with our findings by indicating that ocular discontinuities have the potential to automatically capture our attention. In her studies, Zhaoping compared ocular singletons to orientation singletons and focused on the role of our primary visual cortex during the creation of bottom-up saliency maps for attention guidance.

Dichoptic presentation was extensively studied regarding luminance variation, i.e., exposing an object with different brightness to the left and right eye. How we perceive such a phenomenon was described in the work by de Weert et al.~\cite{de1974binocular} and Teller et al.~\cite{teller1967brightnesses}. Anstis et al.~\cite{anstis1998nonlinear} explored luminance effects in a more diversified way by adding other presentation techniques such as flicker to their comparisons. The research by Formankiewicz et al.~\cite{formankiewicz2009psychophysics} also contributed to that field by examining the ways how such luminance disparities are detected and revealed similarities to the detection of surface properties.

The use of dichoptic presentation for creating novel visual experiences has gained little attention in science. Most closely related to our proposed method is the work by Zhang et al.~\cite{Zhang:2012:BSE}. The authors briefly discuss several ideas for ``Unconventional Binocular Presentation'', with highlighting as one of those proposed applications. Unfortunately, the short note does not evaluate any of those thoughts in more detail. Later, Zhang~\cite{Zhang:2014:SBE} revisited some of the ideas and focused on the luster effects.

Apart from rivalry, dichoptic presentation also takes place in the case of binocular disparity. The term is associated with our ability to extract depth information based on the fact that our eyes have a slight horizontal offset and perceive two slightly different images. For further details on the basics of our stereo vision, we refer to the work of Julesz et al.~\cite{julesz1960binocular, julesz1971foundations} and Marr et al.~\cite{marr1976cooperative, marr1979computational}. Caziot and Backus\cite{Caziot:2015:SOM} performed a thorough study on the effects and parameters of stereoscopic offsets to improve object recognition. We, however, avoid spatial offsets as the position of a data point cannot be changed in general without changing its value, e.g., in plots or graphs. For an overview and classification of the numerous other uses of stereoscopic 3D, we refer the reader to the formalization paper by Schild et al.~\cite{Schild:2015:FPS} and Schild's dissertation~\cite{Schild:2014:DGC}. 

In contrast to those rather common depth perception approaches, the so-called da Vinci stereopsis~\cite{NAKAYAMA19901811} refers to cases where we are able to extract depth information based on monocular occlusions. We point the readers to the state-of-the-art report by Harris et al.~\cite{harris2009role} for an overview. In addition, the more recent findings by Tsirlin et al.~\cite{tsirlin2012vinci} indicate that depth perception in such cases is most likely due to occlusion geometry.

\section{The Deadeye Effect}

We make use of dichoptic presentation and render the target object for one eye only. The other eye sees the plain background at the target location. In contrast to most binocular rivalry experiments, we do not show different objects at the target location. We simply create one image with and one image without the object. Hence, we claim that the conflict produced by the monocular neurons can be easily resolved. Our high-level visual system does not have to decide between two different objects. Instead, we get an image that is not unusual in our daily life. Think of a distant object that you look at through a small hole -- the visual system does not report any conflicts, although the object is seen with one eye only (the dominant one). 

Despite that naturalness, we can recognize the object enhanced with Deadeye in a split second. The object is best described as ``eye-catching'' or ``somehow wrong'', although we can focus on it and perceive all details without trouble. We attribute that popping out to our stereo vision ability. Extracting depth information and fusing two slightly offset images happens nearly instantly. However, depth calculations for the target object result in an error, i.e., that a single scene element cannot be put in a depth relative to other objects and the visual system preattentively recognizes that something went wrong.

An additional explanation to be considered is self-preservation mechanisms: We instantly react to a visual stimulus that is placed right in front of our eye, e.g., by closing our lid and moving back from the possible danger. In such cases, the stimulus is too near and visible with one eye only. Hence, we encourage more in-depth research including brain monitoring via fMRI to determine the exact cause of the observed effect. Our contribution focuses in the first place on establishing Deadeye as a preattentive visualization technique.

\subsection{Enhancing Visualizations With Deadeye}

Though Deadeye can be applied to a broad variety of visualization purposes, we cover a few basic examples to motivate the research on that technique. One possible scenario is the visualization of high-dimensional data via line charts where we want to attract the observers' attention to one or multiple lines of interest (cf. \FG{fig:linechart}). 

\begin{figure}[t!]
\centering
\includegraphics[width=1.0\columnwidth]{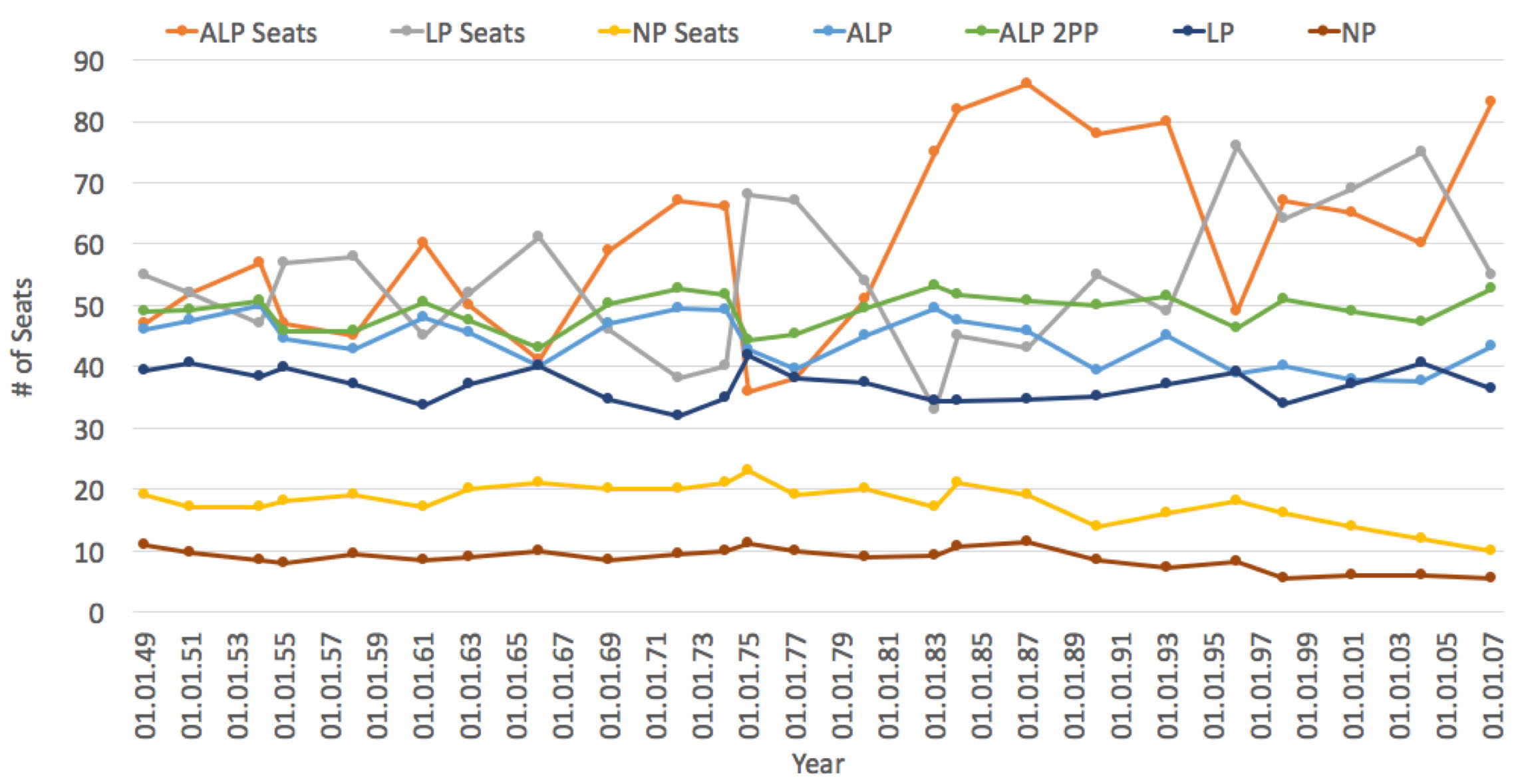}
\caption{An extract from the data about elections to the Australian House of Representatives, 1949-2007, represented as a line chart. In three trials of our prestudy, we enhanced one or multiple lines using Deadeye and asked the subjects to name the corresponding data.}
\label{fig:linechart}
\end{figure}

Applying Deadeye offers a unique advantage that cannot be achieved with any of the existing popout cues: highlighting the line without ``wasting'' any visual dimensions such as color or motion. The technique can be regarded as an additional degree of freedom for visual encoding that is orthogonal to the existing methods. This is essential, as real-world data visualizations tend to utilize as many visual properties (color, stroke width, etc.) as available to cover all data dimensions. Deadeye-enhanced line charts do not have to reserve a visual attribute for highlighting and, thus, can include more data dimensions. 

The same argumentation is also valid for other representations such as bar charts, pie charts, or spider diagrams. Note that applying the ``default'' stereo vision to make a line to stand out from the chart is often not an option for such cases, as depth alters the coordinates of the underlying data and might result in wrong interpretations.

Another mentionable area is comparative visualization. Hereby, one of the common methods is to expose side-by-side views of the data. Such comparison tasks can be quite difficult and are even used as a challenge in so-called spot-the-difference puzzles (cf. \FG{fig:teaser}).

\begin{figure}[t!]
\centering
\includegraphics[width=1.0\columnwidth]{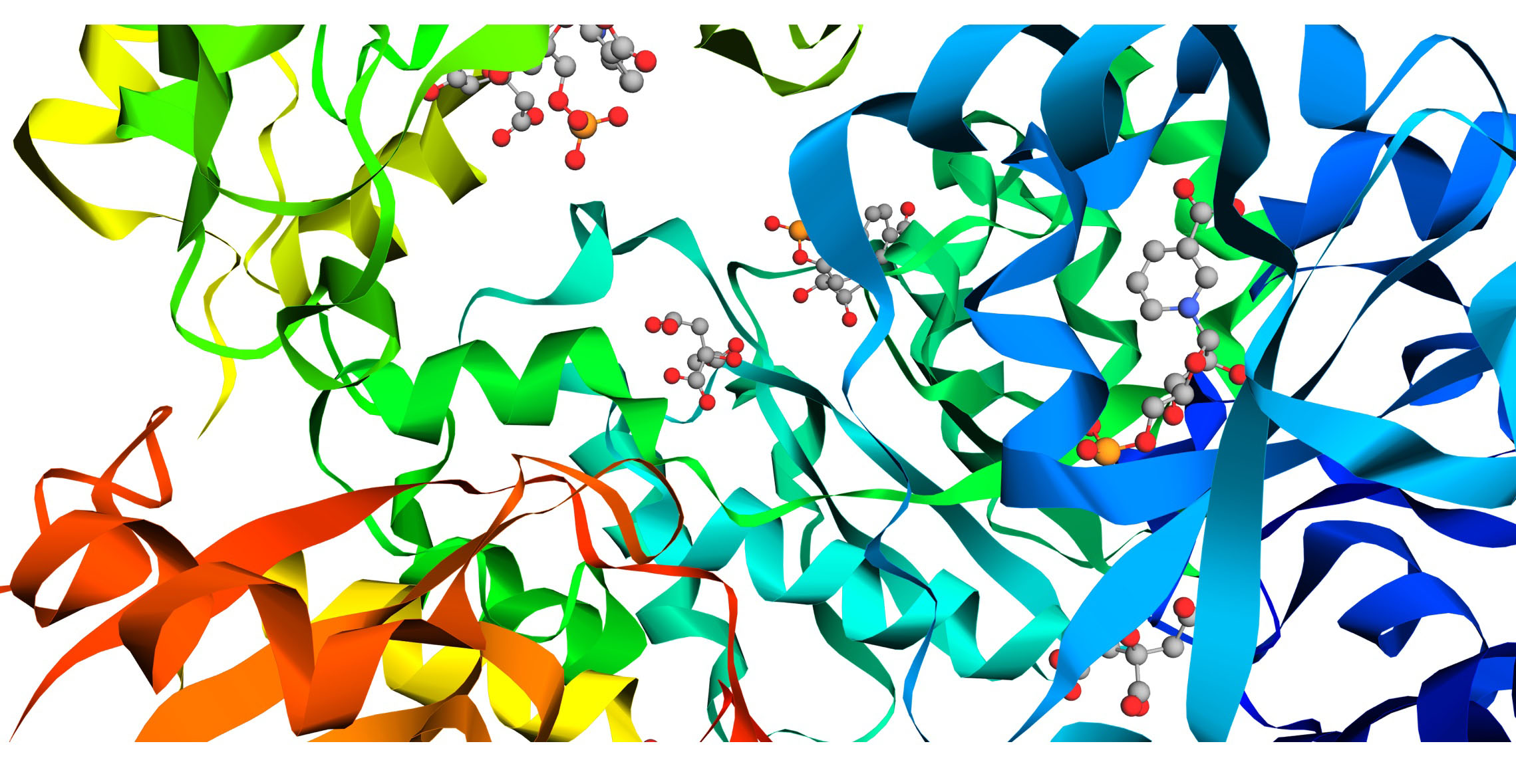}
\caption{Visualization of a transferase (Nicotinic Acid Mononucleotide Adenylyltransferase). As part of our exploratory prestudy, we applied Deadeye on a small number of atoms (between 4 and 8) and asked the participants to count the elements that pop out.}
\label{fig:atoms}
\end{figure}

\begin{figure*}[t!]
\centering
\includegraphics[width=2.1\columnwidth]{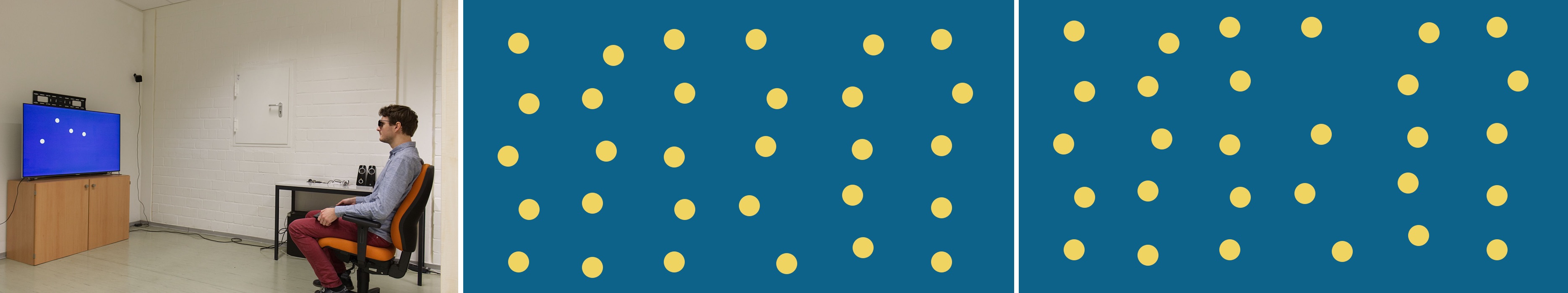}
\caption{Our experimental setup. We equipped participants with active shutter glasses and presented a series of images on a 3D TV. After each image, the participants decided whether there was a target object or not. The images contained a varying number of circles jittered on a 5x6 grid. The two screenshots are a stereo pair example from our largest set with 30 objects. Readers could try to stereo fuse the images to perceive the popout effect.}
\label{fig:study}
\end{figure*}

Our research on Deadeye demonstrates that there could be a much more effective way: the two images can be jointly exposed to our left and right eye. Hereby, the Deadeye effect allows a rapid recognition and localization of missing or differing objects, be it for scientific or entertaining purposes.

Note that though Deadeye requires stereo equipment, the corresponding hardware became commodity in recent years. 3D glasses, stereo projectors, and 3D TVs are wide spread but rarely used in day-to-day visualization applications. Hence, exploiting the devices for information highlighting is an alternative way of ``recycling'' such barely used equipment.

\subsection{Prestudies}

Prior to our main study, we conducted three exploratory experiments regarding the general applicability of Deadeye in different visualization scenarios. Our goals were twofold. First, we verified that subjects are able to exactly locate the popout targets and not only perceive a certain visual mismatch. Second, we determined whether the presence of other cues, such as color, are limiting the applicability of our technique.

We executed the following three prestudies: naming the Deadeye-enhanced line in a line chart (seven subjects, three trials, task cf. \FG{fig:linechart}), counting popping out atoms in a 3D visualization (four subjects, two trials, task cf. \FG{fig:atoms}), and naming the eye-catching image elements in a spot-the-difference puzzle scenario (five subjects, one trial, task cf. \FG{fig:teaser}). We did not impose any time constraints and only recorded the correctness of the answers.

In all cases, participants were able to locate and report the targets without any side effects such as headache. Additionally, in the line chart case, we verified that subjects were able to read the y-axis values aloud for several data points to make sure that information extraction is not affected. We conclude that Deadeye-enhanced objects can indeed be spotted and followed in a variety of visualization scenarios. Moreover, the technique was not affected by the presence of different colors or in a 3D-context (atom counting). Note, however, that the exact interplay of the technique with other preattentive cues is subject of future research.

\section{Evaluation}

We conducted a user study to examine whether Deadeye can be perceived preattentively. Our experiment design is based on existing best practices for determining such effects. The participants are exposed to series of images composed of distractors and possibly a target object as shown in \FG{fig:study}. The participants have to decide whether the target object is present or not. As described in the Related Work section, two approaches exist. One either displays the images for a fixed amount of time (100-250 ms) and measures the error rate, or the image is shown until the participants make a decision. In that case, one also has to consider the reaction time. In our case, we applied the fixed time option and displayed the image for 250 ms.

To prove the preattentive nature of Deadeye, the setup has to be repeated with varying set sizes. The error rate must remain nearly constant, no matter how many distractor objects are presented. Therefore, we considered four sets with the following sizes: 4, 8, 16, and 30. Additionally, we evaluated the conjunction search property of Deadeye by combining our technique with color. Since the common 3D depth is a visual cue that can be combined in parallel with other variables, one might assume the same property for our technique.

\subsection{Preattentive Experiment}

\subsubsection{Hypotheses}

Our main hypothesis is \textbf{H1} is that \textit{Deadeye is indeed a popout technique that can be perceived preattentively.} Therefore, the error rate has to be sufficiently low and remain constant among the different set sizes. Most of the related works consider an error rate below $10 \%$ as optimal. 

Readers might suppose that Deadeye feels uncomfortable for our visual system. Therefore, we formulate a second hypothesis \textbf{H2} as follows: \textit{Deadeye does not lead to headache or other physical strain.} Otherwise, the application possibilities of that visual cue would be rather limited.

\subsubsection{Procedure and Applied Measures}

The study took place in a virtual reality laboratory at our university. After informing participants about the study's procedure, we administered a first questionnaire to assess the general demographic data. In addition, we asked whether the participants have any visual impairment and conducted a hole-in-the-card-test for eye dominance (Dolman method, e.g., \cite{cheng2004association, porac1976dominant}). Participants had to hold a DIN A4 sheet of paper at arm's length and fixate a distant object through a hole in the middle of the paper. They closed the left and right eye in turn and reported whether they still saw the object. If the object disappeared, the closed eye was marked as the dominant eye.


The main part of our study took place in front of a 3D TV (W / H / D 122,40 x 74,10 x 30,60 cm, 1080p) with active shutter glasses and a refresh rate of 60 Hz. The room light was switched off and the curtains were shut in order to minimize the flickering of the shutter glasses. The participants were placed on a chair 280 cm in front of the screen, as depicted in \FG{fig:study}. The distance resulted in a horizontal viewing angle of 12.63$^{\circ}$ from the focus point (vertical: 7.54$^{\circ}$). We gave the participants a keyboard with explicitly marked \textit{yes} and \textit{no} keys and told them to use their thumbs or index fingers for executing the input. The keys were chosen to have the maximum possible distance.

We told the participants that a series of images would be presented. Each image has a blue background and contains a number of yellow circles. Possibly, one of the circles pops out. The image remains visible for a split second. After the inspection, the participants have to press \textit{yes} if they think that the target object was present, and \textit{no} otherwise.

\begin{figure*}[t!]
\centering
\includegraphics[width=2.1\columnwidth]{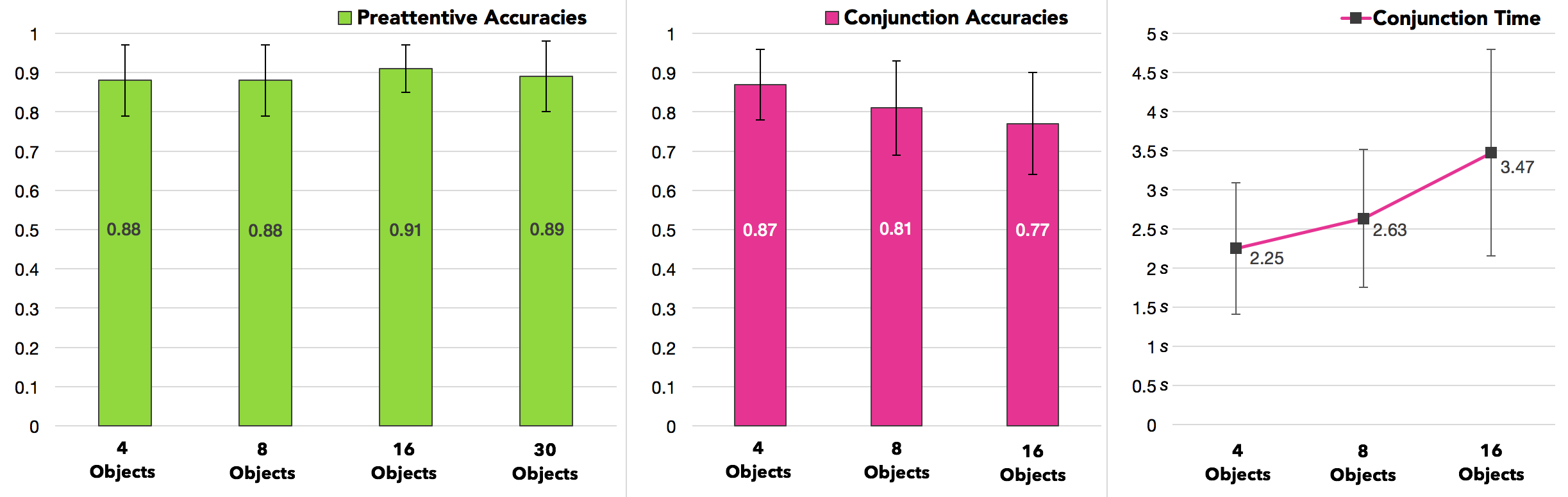}
\caption{The average accuracies for the first experiment are nearly constant and confirm that Deadeye is a preattentive cue. As often the case for such features, false negatives ($M = 0.68$) significantly dominate over false positives ($M = 0.32$). In case of the conjunction search, the images were shown until the participants made a decision. The average accuracies and reaction times significantly decrease with an increasing number of distractors, exposing the serial character of the search.}
\label{fig:accuracy}
\end{figure*}

Between the images, a white crosshair on the same blue background was displayed in the middle. We explicitly advised the participants to focus on the crosshair. This is important as it prevents saccadic eye movements, i.e., the participants' eyes remain in the focus stage when they see the image, because the saccade has an initiation time of approximately 200-250 ms. The crosshair image was always presented for 2500 ms and disappeared when the screen switched to the actual image. We also informed the participants that they did not have to rush with their answer. During that period, the same blue background without the crosshair was exposed.

In addition, we explained that there would be a training stage before each set and that an audio feedback would indicate whether the given answer was correct or not. During the real experiment, the two sounds would be replaced with a third, rather neutral sound. This sound would prevent participants from becoming distracted by thinking about wrong answers and, thus, making subsequent errors due to the lack of concentration. 

Overall, that part of the experiment consisted of four set sizes: 4, 8, 16, and 30 circles. For each set, participants had the chance to practice until they felt comfortable and told the examiner to begin with the real test. After each set, the participants were asked if they needed a pause or wanted to continue. Each set consisted of 48 images, half of them with a target object in a randomized order. Each participant experienced the same configuration. For the 24 images with a target object, 12 were rendered for the right eye and 12 for the left eye. The order was again randomly chosen. We applied that variation mainly to see whether there is a dependency between the error rate for an eye regarding its dominance.

The positions for the circles were randomly generated on a 5x6 grid with a jittering/offset function, as can be seen in \FG{fig:study}. We also left a vertical margin of 11,48 cm and a horizontal margin of 17,44 cm, limiting the overall horizontal viewing angle to about 8.88$^{\circ}$ from the focus point (vertical: 5.22$^{\circ}$). Each circle had a size of 4,59 cm or approximately 0.94$^{\circ}$. This setup led to the image being located in the focal, paracentral, and near-peripheral vision areas.

After the four sets were completed, we administered a web-based effort questionnaire mainly based on the NASA-TLX survey~\cite{hart1988development}. The main reason for choosing NASA-TLX is that the work by Gutwin et al.~\cite{Gutwin:2017:PPI:3025453.3025984} used the same questionnaire for a variety of preattentive cues. Hence, we strived to create a meaningful comparison to other techniques.

The NASA-TLX survey contains six subscales, each scale ranges from 0 to 100 in increments of 5. Following aspects are measured: mental demand (low/high), physical demand (low/high), temporal demand (low/high), performance (good/poor), effort (low/high), and frustration level (low/high). 

We included three additional questions on a seven-point Likert scale ranging from 0 to 6 with larger numbers indicating a more positive outcome: \textit{how well have you perceived the popout object?}, \textit{how sure were you that you made the right decisions?}, and \textit{how well were you able to focus the crosshair?}. We will relate to these custom questions as \textit{clearness}, \textit{decision-making}, and \textit{focus}, respectively. We also included the binary question about \textit{whether the participants experienced any headache or related physical strains}.

\begin{figure*}[t!]
\centering
\includegraphics[width=2.1\columnwidth]{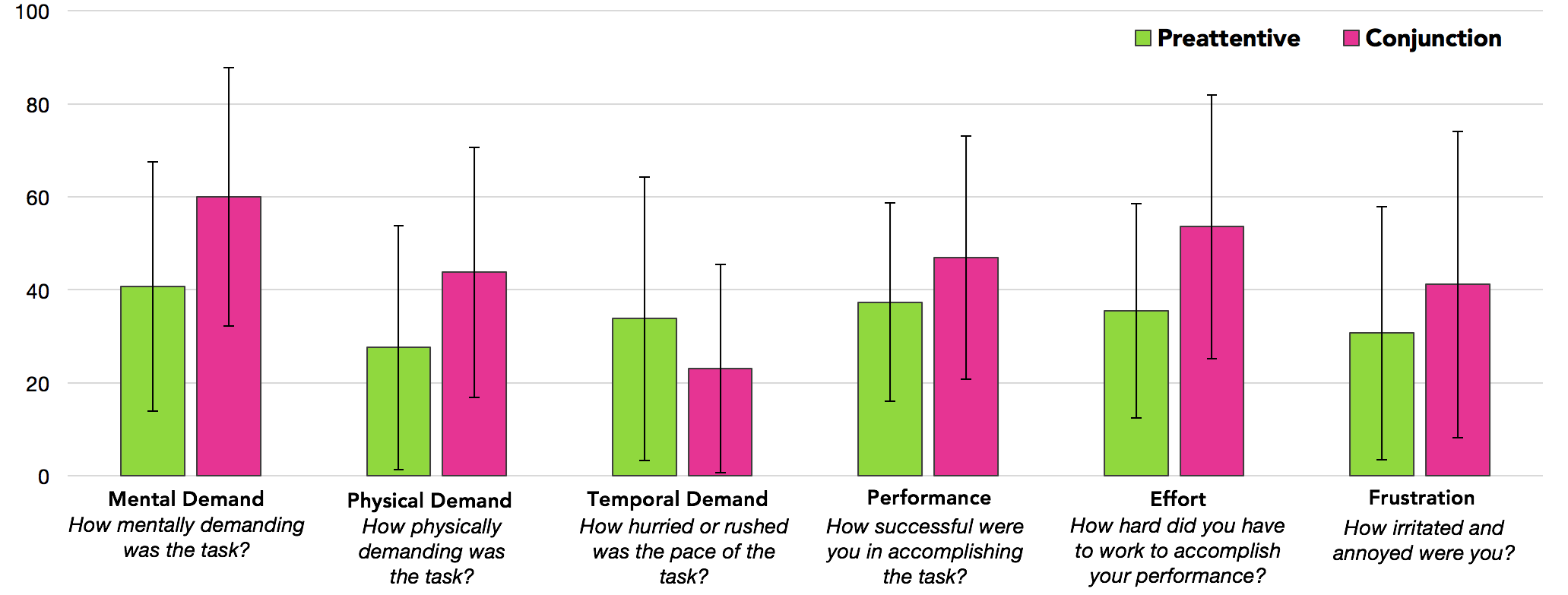}
\caption{Results of the NASA-TLX survey for both experiments. Lower values are preferable.}
\label{fig:nasa}
\end{figure*}

\subsubsection{Performance Results}

In sum, 21 persons (9 females, 12 males), aged 18 to 42 ($M = 25.29,\,SD = 6.21$), participated in the study. Most of them were students ($N = 15$) or employees ($N = 5$). All participants reported normal or corrected to normal acuity and no defects of vision. Most participants had a right dominant eye ($N = 16$), only few had a left dominant eye ($N = 5$).

All performance-based computations were based on detailed automated logging of our experiment application. We measured the accuracy rate for each set. All variables were normally distributed according to Kolmogorov-Smirnov tests. 

The results are depicted in \FG{fig:accuracy}. The average accuracies (4 objects: $M = 0.88,\,SD = 0.09$; 8 objects: $M = 0.88,\,SD = 0.09$; 16 objects: $M = 0.91,\,SD = 0.06$; 30 objects: $M = 0.89,\,SD = 0.09$) are similar to other preattentive visual cues. We applied the repeated measures ANOVA with the set size as within-subject variable to investigate whether the number of distractors influences the accuracy. The result, $F(3,18) = 1.54,\,p = .213$, shows no significant difference in accuracy between the sets, i.e., the number of distractors has no influence on accuracy.

We took a closer look at the wrong answers regarding false negatives, i.e., undiscovered Deadeye-enhanced targets, and false positives. The measured ratio indicates that false negatives ($M = 0.68,\,SD = 0.23$) dominate over false positives ($M = 0.32,\,SD = 0.23$). A paired-samples t-test ($t(20) = -3.66, p = .002$) underlines the significance of that finding.

As we have the detailed data for each image trial of each participant, we also investigated the accuracies based on the object location. The resulting accuracy matrix is presented in \FG{fig:matrix}. Hereby, we only considered images with Deadeye-enhanced objects. Each cell of the 5x6 grid contains the successful recognition rate of a target object at that screen position. We applied a color encoding (red means lower accuracy) to facilitate the visual analysis. The findings indicate that target objects further away from the focus point were harder to recognize.

In addition, we evaluated the relation between the dominant eye of the subject and the target eye of our technique. We had only 5 participants with a left dominant eye, hence we only considered subjects with the right dominant eye for that analysis. We compared the accuracies when the target object was exposed to the dominant eye ($M = 0.84,\,SD = 0.13$) and to the non-dominant eye ($M = 0.86,\,SD = 0.10$). A paired-samples t-test ($t(15) = -1.22, p = .242$) did not expose any significant differences.

\subsubsection{Questionnaire Evaluation}

After the preattentive experiment, we administered a questionnaire based on the NASA-TLX survey (see \FG{fig:nasa}) and our four custom questions. All variables were normally distributed according to Kolmogorov-Smirnov tests. 


The NASA-TLX results for that experiment are rather homogeneous: mental demand ($M = 40.71,\,SD = 26.80$), physical demand ($M = 27.62,\,SD = 26.20$), temporal demand ($M = 33.81,\,SD = 30.41$), performance ($M = 37.38,\,SD = 21.37$), effort ($M = 35.48,\,SD = 23.07$), and frustration ($M = 30.71,\,SD = 27.17$) were rated quite similar. However, the replies were wide spread and the reported min/max values contained both 0 and 100 for each subscale. This is also reflected in the rather large standard deviation.

The custom questions regarding clearness ($M = 4.10,\,SD = 0.99$) and focus ($M = 4.14,\,SD = 1.49$) are rather above average, whereas decision-making ($M = 3.10,\,SD = 1.53$) is slightly below, i.e., the participants were not very sure whether they made the correct decisions. All participants gave a negative answer to the question regarding headache or similar strains.

%
%
%

\subsection{Conjunction Experiment}

An important question for preattentive processing is the possibility of combining multiple features. In their manifold research Triesman et al.~\cite{treisman1980feature, treisman1988feature, treisman1986illusory} pointed out that searching for a single visual stimulus can be conducted in \textit{parallel}, whereas we can perform only a \textit{serial} search for an item defined by the conjunction of two visual variables. An example conjunction of hue and form is depicted in \FG{fig:popout}: the target object could be either a red circle among blue circles or a blue square among red squares. Wolfe et al.~\cite{wolfe1989guided} experimented with different conjunctions of color, motion, size, and orientation, even considering a combination of three cues.


The assumptions of Triesman were partially disproved by McLeod et al.~\cite{mcleod1988visual}, and later, M\"uller et al.~\cite{muller1999visual}: the authors reported that a conjunction of form and motion can be processed in parallel. Furthermore, as pointed out by Townsend~\cite{townsend1990serial}, the general question whether a vision process is serial or parallel is still widely discussed.

Our second experiment was inspired by the findings on a conjunction search by Nakayama et al.~\cite{nakayama1986serial}. The authors discovered that depth can be combined in parallel with other variables such as color and motion. The conducted experiments grouped objects in a front and a back plane. In the case of color, objects in the front plane were red and objects in the back plane were blue. Hence, a target object was either a blue one in the front or a red one in the back, as depicted in \FG{fig:popout}. The authors claim that our visual system processes the planes in an alternating way, and that the search in each plane is executed in parallel. Since stereo vision is also based on binocular disparities, one might assume that our technique could lead to similar effects with regard to a conjunction search. Therefore, we conduct a similar experiment by combining Deadeye with color to evaluate that assumption.

\subsubsection{Hypotheses}

Our technique does not produce 3D depth, nevertheless Deadeye also relies on a binocular disparity. Hence, the similarity in the internal processing, i.e., the merging of two images, might lead to a similar property for our method. Therefore, we suggest that \textit{Deadeye can be combined with color as a second visual cue in parallel} as our last hypothesis \textbf{H3}.

\subsubsection{Procedure and Applied Measures}


Subsequent to the questionnaire of the preattentive experiment, we explained our conjunction search experiment to the participants: Now, half of the circles would be magenta. All yellow circles are popping out, i.e., with Deadeye applied, all magenta circles are not popping out. A target object would have one of the two properties: either it is a yellow circle that does not pop out, or it is a magenta circle that pops out. The main difference from the first experiment is that the image would be displayed until the participants make a decision. We asked the subjects to perform as quickly and as accurately as possible. The reason to change the timing strategy was its more explorative fashion, as we would gather more information about the behavior of our technique for such a conjunction search.

Since the time was not fixed for this experiment, we decided to limit it to three set sizes: 4, 8, and 16 circles. Again, we offered a training phase for each set size and each set was composed of 48 images. Also, 24 images had a target object, 12 with a popping out magenta circle and 12 with a unmodified yellow circle. In each subgroup, 6 of the target objects were rendered for the left eye and 6 for the right eye. The ordering was randomized in the same fashion as in the first experiment.

After completion of the three sets, we requested the participants fill out the same questionnaire as after the first experiment. Overall, the study took about one hour.

\subsubsection{Performance Results}

For the f experiment, we analyzed the accuracy and the reaction time. All variables were normally distributed according to Kolmogorov-Smirnov tests. The average accuracies decrease with the increased object count (4 objects: $M = 0.87,\,SD = 0.09$; 8 objects: $M = 0.81,\,SD = 0.12$; 16 objects: $M = 0.77,\,SD = 0.13$). Our ANOVA result exposes a significant difference in accuracy between the sets ($F(2,19) = 6.81,\,p = .003$). Post hoc Bonferroni tests reveal the details: there is a significant difference in accuracy between 4 and 16 objects ($p = .020$) and no significant difference between 4 and 8 objects ($p = .053$) or between 8 and 16 objects ($p = .352$).

In addition to accuracy, we also measured the reaction time, i.e., the time how long the image was shown until the participants made a decision. A summary is presented in \FG{fig:accuracy}. The reaction time exposes similar tendencies as it increases for larger sets (4 objects: $M = 2.25~s,\,SD = 0.84$; 8 objects: $M = 2.63~s,\,SD = 0.88$; 16 objects: $M = 3.47~s,\,SD = 1.32$). The ANOVA outcome ($F(2,19) = 15.03,\,p < .000$) underlines the significant difference in reaction time between the sets. In detail, the post hoc Bonferroni tests show a significant difference between 4 and 16 objects ($p = .002$) and between 8 and 16 objects ($p < .000$), whereas no significant increase of the reaction time can be observed between 4 and 8 objects ($p < .292$).

We also compared the average accuracies for the popping out magenta target ($M = 0.74,\,SD = 0.24$) and the non-modified yellow target ($M = 0.77,\,SD = 0.10$). The result of a paired-samples t-test ($t(20) = 0.87, p = .393$) did not reveal any significant differences.

\subsubsection{Questionnaire Evaluation}

Again, we administered a questionnaire based on the NASA-TLX survey (see \FG{fig:nasa}) and our four custom questions. All variables were normally distributed according to Kolmogorov-Smirnov tests. 


The NASA-TLX survey exposes rather high scores on the subscales mental demand ($M = 60.00,\,SD = 27.84$) and effort ($M = 53.57,\,SD = 28.29$). In contrast, participants reported to be less hurried according to the temporal demand scale ($M = 23.10,\,SD = 22.33$). The remaining subscales physical demand ($M = 43.81,\,SD = 26.88$), performance ($M = 46.90,\,SD = 26.15$), and frustration ($M = 41.19,\,SD = 32.90$) reported similar outcomes to the first experiment. The reports were also wide spread and contained both 0 and 100 as min/max values for each subscale.

Our custom requests about clearness ($M = 3.00,\,SD = 1.41$), focus ($M = 3.43,\,SD = 1.54$) and especially decision-making ($M = 2.71,\,SD = 1.65$) 
were all rather below the average. Nevertheless, similar to the first experiment, no participant reported a headache or similar strains.

We conducted a paired-samples t-test to compare the outcomes of the questionnaires for the two experiments. There is a significant difference for the following subscales: mental demand ($t(20) = -3.40, p = .003$), physical demand ($t(20) = -2.43, p = .024$), performance ($t(20) = -3.41, p = .003$), focus ($t(20) = 2.37, p = .028$), and clearness ($t(20) = 3.86, p = .003$). In all cases, the conjunction search experiment performed significantly worse compared to the first experiment, as can be seen in \FG{fig:nasa}.

%
%
%

\section{Discussion}

The performance evaluation of the first experiment supports our main hypothesis \textbf{H1} that Deadeye is indeed a preattentive technique. The participants recognized the feature in a 250 ms time frame with an average accuracy of {$\sim90$ \%} independently of the number of distractors. The accuracy is similar to the outcome of most preattentive experiments described in the Related Work section and is considered as a completely sufficient proof. 

With regard to wrong answers, $\sim70$ \% of the errors were false negatives. In other words, it is more likely to overlook a popout target rather than wrongly imagine its presence, which is also a common behavior for preattentive techniques. 

We can also confirm our second hypothesis \textbf{H2} that our technique does not lead to physical strains such as headache. All participants replied with \textit{no} to the corresponding question, both for the preattentive and for the conjunction experiments.

Hence, our finding is indeed an important contribution to the fundamental research, since Deadeye offers several advantages over other preattentive techniques. It is simple to implement (the object has to be removed for one eye), and, in comparison to most other methods, it does not alter any visible properties of the object such as color, position, size, or motion.

Our third hypothesis \textbf{H3} stated that Deadeye can be processed in parallel when combined with color as a second visual cue. Clearly, this is not the case and the hypothesis has to be rejected. The accuracy and the reaction time are both significantly worse when the set size increases. The reaction time for only 4 objects is already greater than 2 seconds (i.e., not preattentive) and increases up to 3.5 seconds for 16 objects. Similarly, such an addition of 12 distractors results in an accuracy drop from 87 \% to 77 \%. Nevertheless, the accuracies show that a conjunction is still possible, although the search has to be executed serially.

Hence, Deadeye does not share similar properties with 3D depth as visual cue regarding conjunction search. Nakayama et al.~\cite{nakayama1986serial} suggest that 3D depth works in parallel because our visual system is able separate the objects in the near and the far plane and to analyze each plane in turn. Hereby, the objects on each plane are processed in parallel. Although Deadeye eliminates any depth information of the target objects and puts them into a zero-depth plane, our experiment resulted in serial processing.

We hypothesize that the distance between the two planes does matter. Deadeye enhancements are rather subtle compared to two planes with a significant depth offset. Hence, we propose to repeat the 3D depth experiment and to gradually reduce the distance between the planes. We assume that there is a certain minimum threshold for parallel processing. If this is not the case, either the two-plane explanation is not fully correct, or our visual system is not able to group Deadeye-enhanced objects into a single depth plane. More aligned with our results, the work by O'Toole et al.~\cite{o1997preattentive} also reported that the question of parallel vs. serial search has several nuances when it comes to 3D depth as a cue. In particular, it does matter whether targets and distractors have crossed or uncrossed disparities, and whether targets are behind or in front of distractors.

\begin{figure}[t]
\centering
\includegraphics[width=0.98\columnwidth]{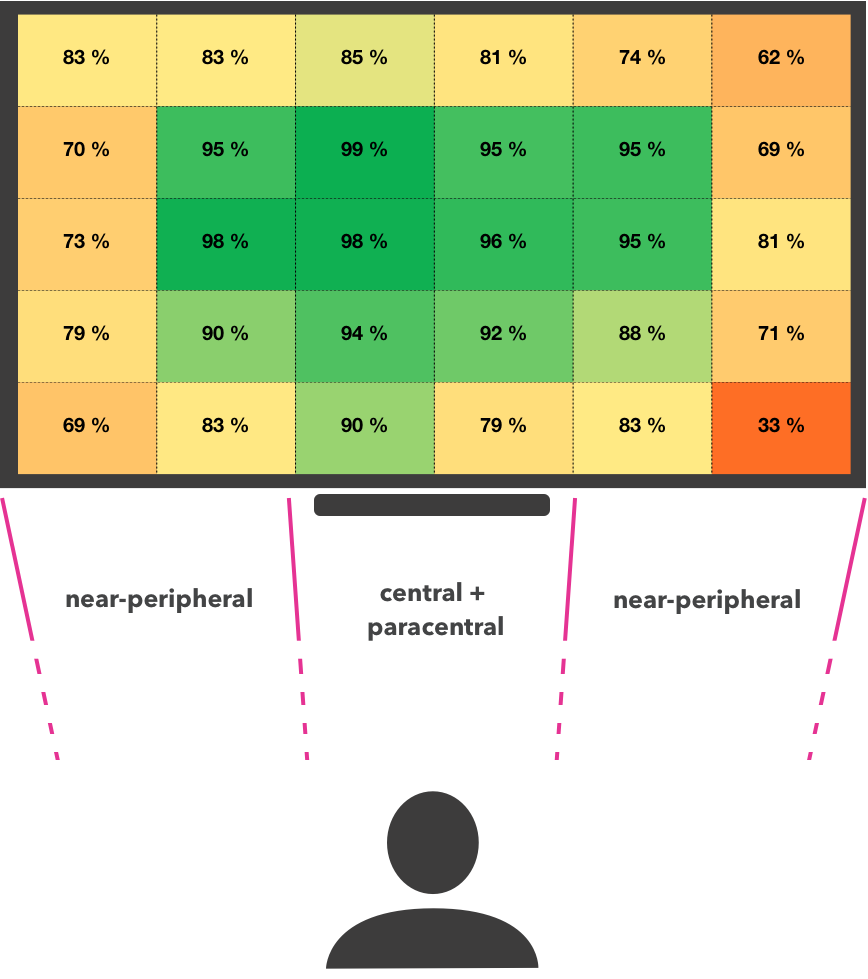}
\caption{Average success rates for the recognition of a Deadeye-enhanced object at each screen position. The matrix indicates an increase of the error rate with increasing distance from the focus point.}
\label{fig:matrix}
\end{figure}

To conclude, if a conjunction search is required, the 3D depth has an advantage. However, preattentive cues are rarely used for conjunctions. For the majority of popout applications, 3D depth has the drawback on modifying the position of the target object, which is often a significant issue. Deadeye, in contrast, preserves the correct object position.

An interesting aspect for preattentive cues is the subjective user perception. The first thing we notice is that users tend to underestimate their performance. This can be seen in the rather average results on the performance scale of the NASA-TLX (\textit{how successful do you think you were in accomplishing the goals?}) and our custom question on decision-making (\textit{how sure were you that you made the right decisions?}). Both results contradict the achieved accuracy of $\sim90$ \%. On the other hand, this is not surprising when we recall that the preattentive analysis happens unconsciously. The images were shown for a split second, and subjects had to rely on a rather unchecked, intuitive result provided by the visual system. 

Our second experiment was rated significantly worse regarding mental and physical demand, performance, and clearness. Overall, we can conclude that the task was more challenging and less straightforward compared to the first experiment. The target object could be only discovered by a conscious and serial search, which explains these scores. The participants also reported a significant decrease in their ability to focus on the crosshair before each image. We attribute that outcome to the fact that a serial search does not benefit from a stationary period, and an immediate saccadic eye movement appeared more efficient to the most subjects.

When comparing the NASA-TLX results of our first experiment to the outcomes reported by Gutwin et al.~\cite{Gutwin:2017:PPI:3025453.3025984}, we can observe that Deadeye performed slightly better than existing cues (note: be aware of different scales). Especially the subscales performance and effort appear to be significantly better and indicate that Deadeye can compete even against canonical cues such as color. 

However, that comparison should be considered with a grain of salt. The authors explored the cues with a significantly larger viewing angle to study the behavior regarding the peripheral vision. Hence, the difference in results might be due to the viewing angle. Unfortunately, there is not much work on the comparison of different features, and we encourage to undertake further steps towards a comprehensive overview.

Our analysis shows an additional relation to the peripheral experiments of Gutwin et al., as the authors have discovered that certain cues were less efficient when applied far from the focus point. Similarly, our accuracy matrix in \FG{fig:matrix} exposes a drop-off in performance for the outer columns. This indicates that Deadeye suffers under the same reduced peripheral efficiency as, e.g., shape and color. In our experiment, the image area covered a total visual angle of $\sim19^{\circ}$ and included the central, paracentral, and the near-peripheral regions. Our depth perception decreases with the distance from the focus point (e.g., \cite{mochizuki2012magnitude}) and nearly disappears in the peripheral region. Since we assume that the depth computation process is at least partially responsible for discovering Deadeye-enhanced objects, such a drop-off or even a complete inapplicability in far-peripheral regions appears rather comprehensible.

Another observation from our experiment is that it does not matter whether Deadeye is rendered on the dominant or the non-dominant eye. That behavior is rather in line with the findings of, e.g., Logothetis et al.~\cite{logothetis1996rivalling}, stating that binocular rivalry is not eye-dependent.

Certainly, Deadeye has limitations that need to be discussed. First of all, Deadeye requires a stereoscopic environment, since a different image needs to be exposed to each eye. This fact renders the technique less convenient for everyday usage and requires additional hardware.

In addition, we suggest to evaluate Deadeye in a 3D environment such as virtual reality. Our technique eliminates the depth cues for the target object. Thus, we assume that Deadeye might perform differently in such a scenario. Another disadvantage is that Deadeye is not applicable for one-eyed users. On the other hand, this kind of visual restriction is significantly less frequent compared to, e.g., color vision deficiency.

A mentionable difference between Deadeye and most other preattentive cues is its binary character. Our current method has no graduation, i.e., the object is either popping out or not. In contrast, cues such as hue can be applied in varying intensity levels, which is an additional degree of freedom and, e.g., allows a differentiation between target objects.

\section{Conclusion and Future Work}

Our contribution is \textit{Deadeye}, a novel preattentive visualization technique. Preattentive cues are used in a plethora of visualization approaches and interaction paradigms and allow us to enhance objects of interest such that they pop out independently from the number of distractors. Our technique contributes to the fundamental research in visualization in two ways. First, a discovery of a preattentive cue opens the door for further research and several possible applications. Second, Deadeye is the first preattentive technique that does not alter any visible properties of the target object. In contrast to existing cues, our method does not displace, recolor, reshape or animate the target. Hence, the probability of misinterpreting the data is minimized, which is a significant benefit compared to existing methods.

Deadeye is based on presenting two slightly different images to the human visual system. Hereby, the target object that should pop out is rendered for one eye only. We evaluated the method by a state-of-the-art study being commonly applied for popout variables and demonstrated that Deadeye can indeed be perceived preattentively. Three smaller, explorative experiments illustrate real-world applications of our technique in different visualization setups. In addition, we conducted a conjunction search experiment by combining Deadeye with color as a second cue. Our results showed that, in contrast to common 3D depth, Deadeye conjunctions cannot be processed in parallel.

Our initial findings encourage additional research questions that we will address in future work. Our preliminary lab experiments indicate that Deadeye delivers robust performance even if distractors of different kind are present. Our first tests with objects of different shape and color support that assumption and will be further extended in the future. This would be a significant advantage over many of other preattentive cues, since the data to be visualized is often composed of heterogeneous objects.

Another interesting experiment would be to apply our technique to moving objects to evaluate how the effect performs in dynamic scenes. Furthermore, we suggest integrating and evaluating Deadeye in sophisticated attention guidance setups and existing application-level tools that already rely on popout techniques. In summary, we believe that dichoptic presentation has the potential to become a useful ingredient to the visualization toolbox beyond just stereoscopic 3D.

\acknowledgments{We are immensely grateful to Christine Pickett for her comments that greatly improved the manuscript. We also wish to thank Rolf Rehe for reminding us how we used the wall-eyed vision trick to solve spot-the-difference puzzles when we were kids. We would also like to show our gratitude to the anonymous reviewers for their detailed feedback and all the valuable suggestions they made.}

\bibliographystyle{abbrv-doi}

\bibliography{deadeye}
\end{document}